\documentclass[a4paper,11pt]{article} 
\usepackage{graphicx} 

\newcommand{\del}{\partial} 
\newcommand{\tr}{{\rm tr}}

\makeatletter
\newtoks\@stequation

\def\subequations{\refstepcounter{equation}%
  \edef\@savedequation{\the\c@equation}%
  \@stequation=\expandafter{\theequation}
  \edef\@savedtheequation{\the\@stequation}
  \edef\oldtheequation{\theequation}%
  \setcounter{equation}{0}%
  \def\theequation{\oldtheequation\alph{equation}}}

\def\endsubequations{%
  \ifnum\c@equation < 2 \@warning{Only \the\c@equation\space subequation
    used in equation \@savedequation}\fi
  \setcounter{equation}{\@savedequation}%
  \@stequation=\expandafter{\@savedtheequation}%
  \edef\theequation{\the\@stequation}%
  \global\@ignoretrue}

\def\eqnarray{\stepcounter{equation}\let\@currentlabel\theequation
\global\@eqnswtrue\m@th
\global\@eqcnt\z@\tabskip\@centering\let\\\@eqncr
$$\halign to\displaywidth\bgroup\@eqnsel\hskip\@centering
     $\displaystyle\tabskip\z@{##}$&\global\@eqcnt\@ne
      \hfil$\;{##}\;$\hfil
     &\global\@eqcnt\tw@ $\displaystyle\tabskip\z@{##}$\hfil
   \tabskip\@centering&\llap{##}\tabskip\z@\cr}

\makeatother

\begin{document} 
\thispagestyle{empty} 
\vspace*{-15mm} 
\baselineskip 10pt 
\begin{flushright} 
\begin{tabular}{l} 
{\bf September 2000}\\ 
{\bf hep-th/0009098}\\ 
\end{tabular} 
\end{flushright} 
\baselineskip 24pt 
\vglue 10mm 
\begin{center} 
{\Large\bf 
Perturbative ultraviolet and infrared dynamics of 
       noncommutative quantum field theory 
} 
\renewcommand{\thefootnote}{\alph{footnote}} 
\footnote{Talk given at XXXth International Conference on 
          High Energy Physics (ICHEP2000), July 27-August 2, 
          Osaka, Japan} 
\vspace{8mm} 

\baselineskip 18pt 
\def\thefootnote{\fnsymbol{footnote}} 
\setcounter{footnote}{0} 
{\bf 
 Masashi Hayakawa 
} 
\vspace{5mm} 

{\it Theory Division, KEK, Tsukuba, Ibaraki 305-0801, Japan} 

\vspace{10mm} 
\end{center} 

\begin{center} 
{\bf Abstract}\\[7mm] 
\begin{minipage}{12cm} 
\baselineskip 16pt 
\noindent 
 Perturbative aspects of ultraviolet and 
infrared dynamics of noncommutative quantum field theory 
is examined in detail. 
 It is observed that
high loop momentum contribution to 
the nonplanar diagram develops a new infrared singularity 
with respect to the external momentum. 
 This singular behavior is 
closely related to that of 
ultraviolet divergence of planar diagram. 
 It is also shown that 
such a relation is precise in noncommutative Yang-Mills theory, 
but the same feature does not 
persist in noncommutative generalization of QED. 
\end{minipage} 
\end{center} 
%
\newpage
\baselineskip 18pt 
\section{Introduction} 
\label{sec:intro} 

 The primary purpose here is to observe 
the perturbative aspects of noncommutative quantum field theory. 
 After viewing in Sec.~\ref{sec:motivation} one motivation 
why noncommutative quantum field theory becomes interesting, 
we argue ultraviolet (UV) property derived from 
perturbative consideration \cite{Bigatti,Ishibashi_NCYM} 
with an introduction to perturbative framework of 
noncommutative quantum field theory 
in Sec.~\ref{sec:perturbation}. 
 In Sec.~\ref{sec:IR}, 
we examine the infrared (IR) aspects, 
and show how it is closely related to UV ones, 
especially in noncommutative Yang-Mills (NCYM) theory. 
 Final section is devoted to the discussion and conclusion. 

\section{Motivation} 
\label{sec:motivation} 
 Noncommutative field theory appears in the 
matrix models \cite{Connes,Aoki_NCYM}. 
 The matrix model conjecture \cite{BFSS,IKKT,DVV} 
is intended to 
provide a constructive definition of the superstring theory 
and to extract nonperturbative consequences of 
the interacting superstring dynamics, 
which will enable us to ask whether string theory is real or not. 
 However, for instance, IIB matrix model \cite{IKKT} 
does not have any dimensionless coupling constant 
which can be decreased at will. 
 Therefore, 
the direct perturbative analysis is not available in that model. 

 One way to see appearance of 
noncommutative Yang-Mills theory from matrix model 
is to expand IIB matrix model action around BPS solution 
\cite{Aoki_NCYM} 
\begin{equation} 
 \left[ X^\mu, X^\nu \right]  
  = -i C^{\mu\nu} 1_{N}\, . 
   \label{eq:algebra_X} 
\end{equation} 
where the size of bosonic matrices $X^M$ ($M=1, \cdots, 10$) 
is taken to be infinite. 
 $C^{\mu\nu}$ $=$ $-C^{\nu\mu}$ denotes the abelian part of 
the field strength $F_{\mu\nu}$, 
where $\mu$ is restricted to $1, \cdots, 4$. 
 Reminding that IIB matrix proposes that 
the eigenvalues of $X^M$ constitutes the points 
of the universe, at least semiclassically, 
the above relation (\ref{eq:algebra_X}) implies that 
the location of each point $x^\mu$ is uncertain
in those four directions: 
\begin{equation} 
 \left| x^\mu \right| \left| x^\nu \right| \ge 
  2\pi \left| C^{\mu\nu} \right| 
  \quad {\rm for}\ \mu \ne \nu\, , 
\end{equation} 
and that $C^{\mu\nu}$ characterizes the minimal area 
of accuracy in each two-dimensional plane. 

 IIB matrix model action gives an action 
with respect to the fluctuation $a_\mu$ 
(and the other six bosonic coordinates 
and fermionic variables) around the previous BPS solution 
(\ref{eq:algebra_X}), 
where $X^\mu = X^\mu_{(0)} + C^{\mu\nu} a_\nu$ 
and $X^\mu_{(0)}$ is the classical part satisfying 
eq.~(\ref{eq:algebra_X}). 
 Indeed, through the map called as ``Weyl correspondence'', 
the system can be described in terms of 
a four-dimensional field theory 
(See Ref.~\cite{Aoki_NCYM} on its detail.). 
 The resulting theory is ${\cal N}=4$ supersymmetric 
noncommutative Yang-Mills (NCYM) theory, 
where the fields in the action are multiplied by 
the star-product 
(See, e.g, Ref.~\cite{Zacho} 
for its original geometric construction, 
its appearance in the other physical systems 
and references.) 
defined by 
\begin{eqnarray} 
 && 
 (f*g)(x) = 
  \left. 
   \exp \left( 
          \frac{1}{2i} 
          \del_\mu C^{\mu\nu} \del^\prime_\nu 
        \right) 
   f(x)\, g(x^\prime) 
  \right|_{x^\prime \rightarrow x} \, , 
\end{eqnarray} 
where $C^{\mu\nu}$ is the parameter appearing before. 
 For the coordinates, for instance, we obtain 
\begin{equation} 
 x^\mu * x^\nu - x^\nu * x^\mu = 
  - i C^{\mu\nu}\, . 
\end{equation} 
 This algebraic relation is isomorphic to the original algebra 
(\ref{eq:algebra_X}) satisfied by the background represented 
by matrices. 

 Now the coupling constant $g_{\rm NCYM}$ in 
the resulting Yang-Mills theory is 
given by 
\begin{equation} 
 g^2_{\rm NCYM} = \frac{4\pi^2 g_{\rm IIB}^2}{C^2} \, . 
\end{equation} 
 Here $g_{\rm IIB}$ is the coupling constant 
of the original matrix model and also has dimension 
of the length squared. 
 In the canonical basis of $C^{\mu\nu}$, 
it has been assumed that 
$C^{\mu\nu}$ = $C$ 
( $i\sigma_2 \otimes 1$ + $1 \otimes i\sigma_2$ ) 
for simplicity. 
 Thus, 
by taking $C^{\mu\nu}$ sufficiently large 
compared to $g_{\rm IIB}$, 
we get weak coupling NCYM theory. 
 Hence, we can investigate the dynamics 
of matrix model 
by analyzing the quantum mechanical aspect 
of NCYM theory. 
 The challenge is to show the existence of gravity and string 
in the quantized NCYM system. 
 If this is shown, 
it will give a strong evidence that 
supports the entire matrix model as 
constructive definition of superstring. 
 The structures of deformation of the open string algebra 
due to closed string background 
and the background independence \cite{Seiberg} 
might also be further demonstrated. 
 
 However, since we do not know noncommutative quantum field 
theory itself so well, 
we are inclined to begin with examination of simpler systems, 
and capture the generic aspects possessed by 
noncommutative quantum field theory. 

 In the succeeding sections, 
we would like to see a few remarkable features 
of the perturbative noncommutative field theory. 

\section{Perturbative analysis of noncommutative field theory } 
\label{sec:perturbation} 
 In order to figure out the basic facets of 
the perturbative framework \cite{Pre_NC} of 
noncommutative quantum field theory, 
we pay our attention to the noncommutative extension of 
a real scalar $\phi^4$ theory in Euclidean 
four-dimensional space 
(see Ref.~\cite{Ishibashi_NCYM,MRS} on its detail) 
\begin{eqnarray} 
 && 
 \displaystyle{ 
  S_{\phi^4_4} = 
  \int d^4 x 
  \left[ 
   \frac{1}{2} \del_\mu \phi *\del_\mu \phi 
   + 
   \frac{1}{2} m^2 \phi *\phi 
   + 
   \frac{\lambda}{4} \phi *\phi *\phi *\phi 
  \right] \, . 
 } 
  \label{eq:action} 
\end{eqnarray} 
 The procedure for perturbation theory 
is the same as that in the ordinary field theory. 
 The first task is to derive 
Feynman rule from the action (\ref{eq:action}). 
 Then we apply Feynman rule to write down the diagrams 
relevant to the process 
and to the order of the coupling constant, 
and evaluate the associated contributions. 

 To derive Feynman rule, 
it is convenient to work in momentum space: 
\begin{equation} 
 \phi(x) = 
  \int \frac{d^4 p}{(2\pi)^4} 
  e^{ip\cdot x} \widetilde{\phi}(p)\, . 
\end{equation} 
 Then the star product works on the basis elements 
$ e^{i p\cdot x}$ in such a way that 
\begin{eqnarray} 
 e^{ip\cdot x} * e^{iq\cdot x} &=& 
   e^{ \frac{1}{2i} \del \wedge \del^\prime } 
   e^{ip\cdot x} e^{iq\cdot x^\prime} 
   \nonumber \\ 
 &=& 
  e^{\frac{i}{2} p \wedge q} \, 
  e^{i(p+q) \cdot x} \, , 
\end{eqnarray} 
where $p \wedge q \equiv p_\mu C^{\mu\nu} q_\nu = - q \wedge p$. 
 This extra phase factor is reminiscent of 
noncommutativity of the star product. 

 First, we consider the propagator. 
 Due to the total momentum conservation of the system, 
only one momentum is linearly independent. 
 Thus, there is no room for phase factors to enter; 
the propagator is the same as in the ordinary field theory 
\begin{equation} 
 \left< 
  \widetilde{\phi}(p) \widetilde{\phi}(q) 
 \right> 
 = 
 (2\pi)^4 \delta^4(p + q)\, 
 \frac{1}{p^2 + m^2}\, . 
\end{equation} 

 However, the interaction vertex picks up nontrivial phase factor 
\begin{eqnarray} 
 && 
 \displaystyle{ 
  \int \prod_{j=1}^4 \frac{d^4 p}{(2\pi)^4}\, 
  (2\pi)^4 \delta^4(p_1 + \cdots  + p_4) 
 } 
  \nonumber \\ 
 && \quad 
 \displaystyle{ 
  \times 
  \frac{\lambda}{4} 
  \exp \left( 
        \frac{i}{2} 
        \sum_{i<j} p_i \wedge p_j 
       \right) 
  \widetilde{\phi}(p_1) \cdots \widetilde{\phi}(p_4) 
 } 
  \, , 
  \label{eq:vertex_1} 
\end{eqnarray} 
from the star-product. 
 Due to this phase factor,
the interaction has only cyclic symmetry, 
in contrast to the point vertex 
in the ordinary real scalar $\phi^4$ field theory 
which is invariant under the whole permutation group. 
 Such a loss of symmetry of the vertex 
is better described if the vertex and legs 
get some width.
 The width does not permit us to exchange 
the two neighboring external legs, for instance. 
 Alternatively,
if multiple lines, rather than one, are assigned 
to each leg, 
they also retain only cyclic symmetry. 
 The ordinary Yang-Mills theory is such an example 
\cite{'tHooft:planar}. 
 It gives 
natural description of the propagator 
based on the double line representation, 
each line carrying the color degrees of freedom. 
 In fact, also to the noncommutative field theory, 
the double line representation will turn out 
to be suitable 
 This aspect is most crucial to observe 
the important fact 
that noncommutative field theory 
gives the same UV structure as that 
of the corresponding ordinary large N field theory. 

\begin{figure} [t] 
\begin{center} 
 \includegraphics[scale=0.3]{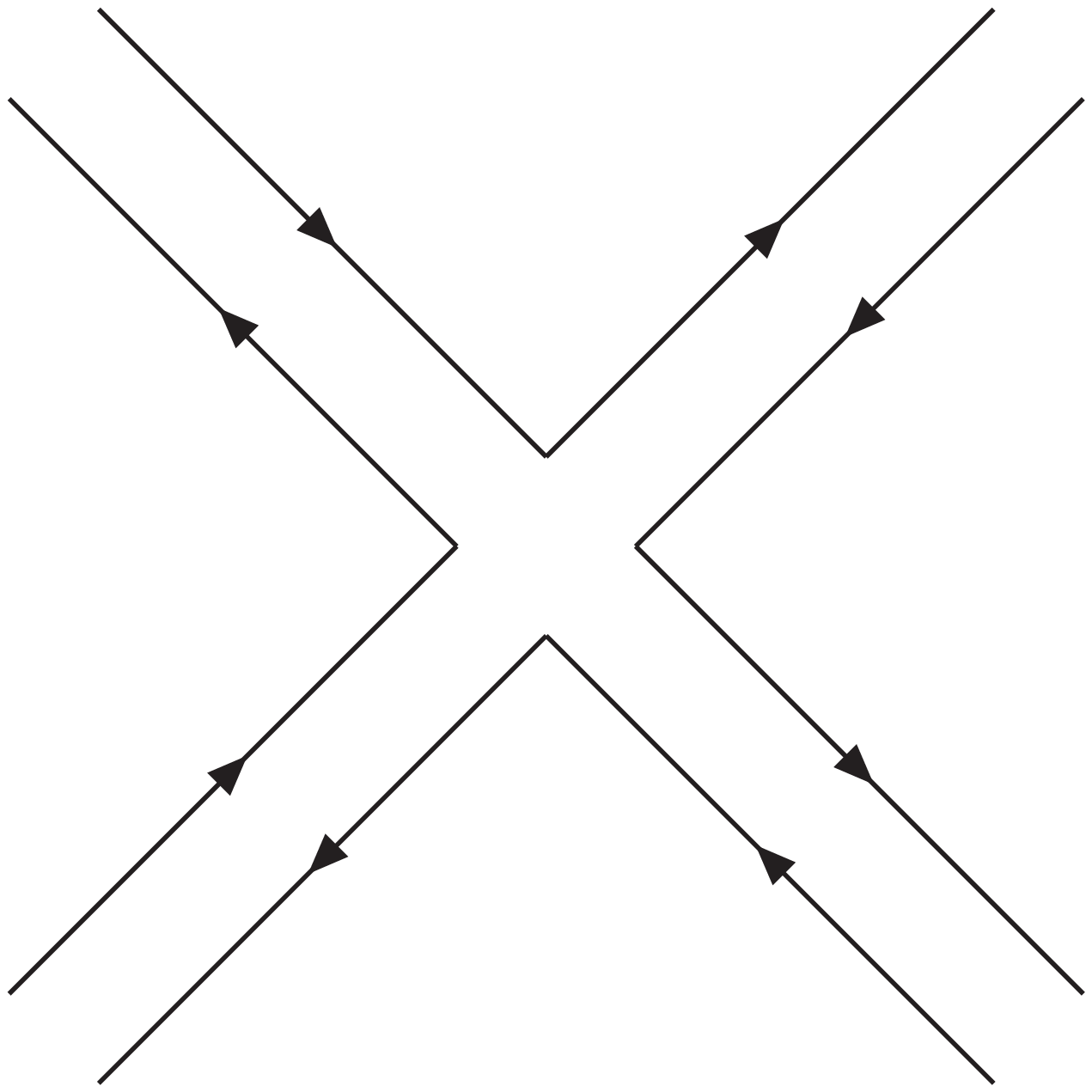} 
 \caption{Double line representation for $\phi^4$-vertex.} 
 \label{fig:double_Feynman} 
\end{center} 
\end{figure}

 To pursue the best picture, 
we attempt to write each momentum $p_j$ as a combination of 
outgoing and incoming momenta 
\begin{equation} 
 p_j = k_j - k_{j-1}\, . 
\end{equation} 
($k_0 = k_4$), 
and examine the consequences. 
 By drawing the flow of each new momentum $k_j$, 
we get the double line representation for the vertex 
as shown in Fig.~\ref{fig:double_Feynman}. 
 To see that this parametrization is natural, 
we rewrite the vertex (\ref{eq:vertex_1}) in terms of $k_i$ 
\begin{eqnarray} 
 && 
 \int 4^4 \prod_{j=1}^4 \frac{d^4 k_j}{(2\pi)^4} 
 (2\pi)^4 \delta^4(k_1 + \cdots + k_4) 
  \nonumber \\ 
 && \qquad \qquad 
  \times 
  \frac{\lambda}{4} 
  \left[ 
   e^{\frac{i}{2} k_4 \wedge k_1} \widetilde{\phi}(k_1 - k_4) 
  \right] 
  \times \cdots 
  \times 
  \left[ 
   e^{\frac{i}{2} k_3 \wedge k_4} \widetilde{\phi}(k_4 - k_3) 
  \right] \, . 
   \label{eq:vertex_NC} 
\end{eqnarray} 
($4^4$ is the Jacobian factor due to the change of the variables 
from $p_i$'s to $k_j$'s.) 
 There, the expression of the phase factor has simplified: 
$\sum_{i<j} p_i \wedge p_j$ $=$ $\sum_{i=1}^4 k_i \wedge k_{i+1}$, 
and each piece has been placed in 
front of the field which depends on the same pair of momenta 
in eq.~(\ref{eq:vertex_NC}). 
 Regarding the quantity in each bracket 
as a matrix element 
\begin{equation} 
 \phi[k_1, k_2] = e^{\frac{i}{2} k_1 \wedge k_2} 
                   \widetilde{\phi}(k_2 - k_1) 
\end{equation} 
labeled by two momenta, it is easy to see that 
it constitutes 
a ``hermitian'' matrix 
\begin{equation} 
 \left( \phi[k_1, k_2] \right)^* = \phi[k_2, k_1]\, , 
\end{equation} 
from reality condition 
$ \widetilde{\phi}(p)^* = \widetilde{\phi}(-p)$ 
in momentum space. 

 What we learn here is that, 
such a hermitian quantity is a building block 
of the interaction vertex 
in the noncommutative real scalar theory, 
and expresses Feynman rule compactly by 
the ``double-line'' representation. 
 
 Taking into account of these facts, 
we are inclined to recall 
the ordinary hermitian matrix field theory 
with quartic interaction 
\begin{eqnarray} 
 && 
 S_{\left[ \Phi_4^4 \right]_N} = 
 \int d^4 x \, 
 \tr 
 \left[ 
  \frac{1}{2} \del_\mu \Phi \del_\mu \Phi 
  + 
  \frac{1}{2} m^2 \Phi^2 
  + 
  \frac{\lambda_H}{4N} \Phi^4 
 \right] \, , 
\end{eqnarray} 
where $\Phi(x)$ is 
an $N \times N$ hermitian matrix-valued field. 
 The factor $1/N$ in front of quartic interaction 
is prepared for the future purpose to take large N limit. 
 In terms of the momentum space variable $\widetilde{\Phi}(p)$, 
the interaction vertex of large N hermitian matrix field theory 
takes the form 
\begin{eqnarray} 
 && 
 \int 4^4 \prod_{j=1}^4 \frac{d^4 k_j}{(2\pi)^4} 
 (2\pi)^4 \delta^4(k_1 + \cdots + k_4) 
  \nonumber \\ 
 && \qquad \quad 
  \times 
  \frac{\lambda_H}{4 N} 
  \widetilde{\Phi}_{i_4}^{\ i_1}(k_1 - k_4) 
  \times \cdots 
  \times 
  \widetilde{\Phi}_{i_3}^{\ i_4}(k_4 - k_3) \, . 
   \label{eq:vertex_large_N}
\end{eqnarray} 
 Comparison of eq.~(\ref{eq:vertex_NC}) 
and eq.~(\ref{eq:vertex_large_N}) shows that 
Feynman diagrams drawn in noncommutative $\phi^4$ theory 
and large N hermitian matrix field theory 
coincide with each other, 
including their combinatoric factors. 

\begin{figure} 
\begin{center} 
 \includegraphics[scale=0.4]{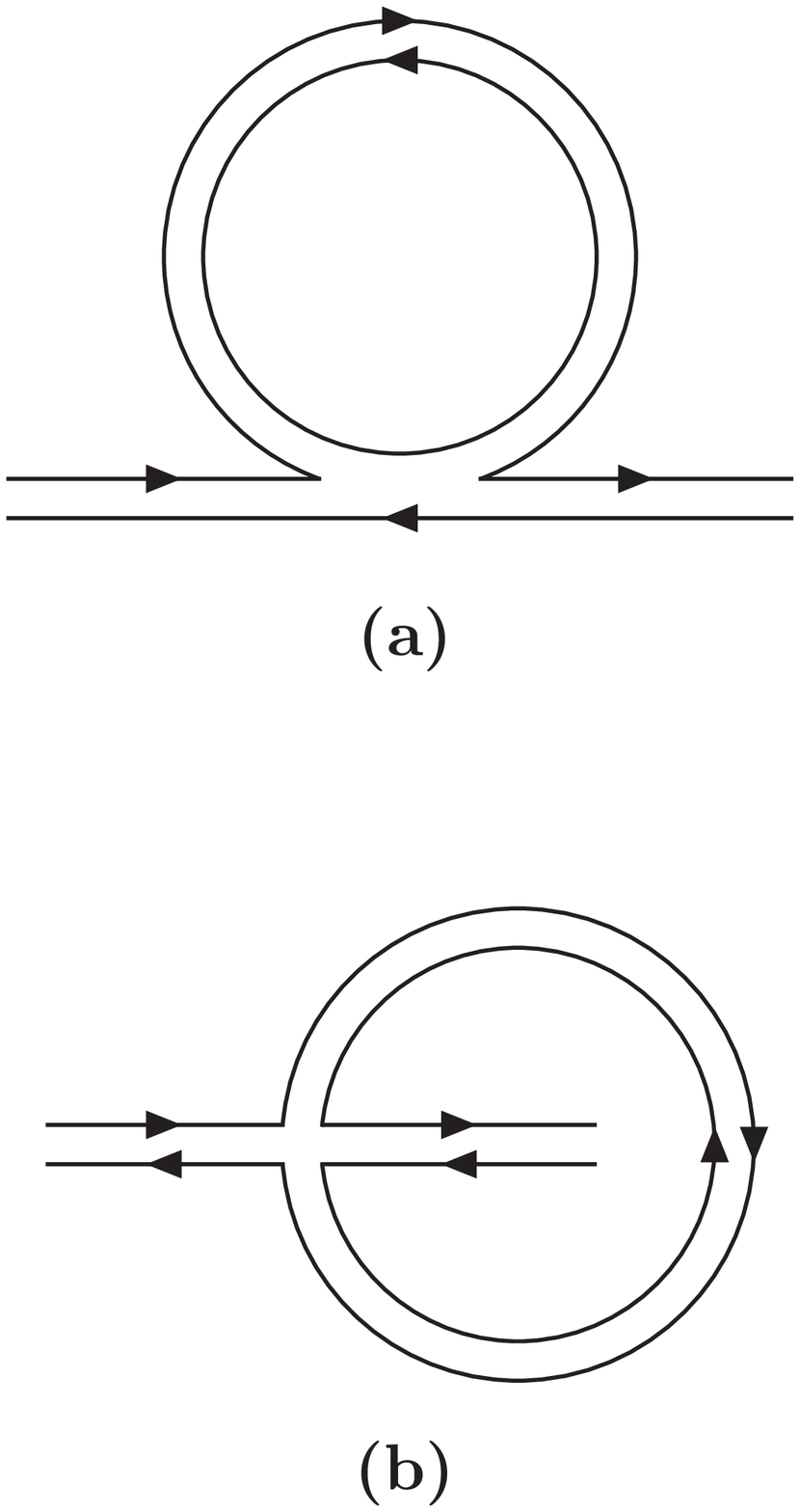} 
 \caption{One-loop correction to two-point functions 
 in $\phi^4$ theory.} 
 \label{fig:two_point} 
\end{center} 
\end{figure}

 Explicit evaluation of the diagrams show that 
the phase factor in noncommutative field theory 
plays the role of the color indices 
carried by the matrix field in the large N field theory; 
the phase factor distinguishes 
planar and nonplanar diagrams. 
 To illustrate this aspect in more detail, 
we consider one loop contribution to 
the two point function in both theories. 
 There are two types of diagrams as shown in 
Fig.~\ref{fig:two_point}. 
  As noted before, in both theories, 
we can draw the same diagrams. 
 Thus, also in the side of noncommutative field theory, 
we can use the same terminology 
to distinguish these two types of the diagrams 
as in the ordinary field theory . 
 That is, Fig.~\ref{fig:two_point}(a) is called as planar 
while Fig.~\ref{fig:two_point}(b) as nonplanar. 

 First we recall the situation 
in the side of large N field theory 
\begin{eqnarray} 
 \Pi^{\rm large\ N}_{\rm planar}(p, \lambda_H) 
  &=& 
    \frac{\lambda_H}{2} 
    \int \frac{d^4 q}{(2\pi)^4} 
    \frac{1}{q^2 + m^2} \, , 
     \nonumber \\ 
  \Pi^{\rm large\ N}_{\rm nonplanar}(p, \lambda_H) 
  &=& 
    \frac{1}{N} 
    \frac{\lambda_H}{4} 
    \int \frac{d^4 q}{(2\pi)^4} 
    \frac{1}{q^2 + m^2} \, . 
\end{eqnarray} 
 Both diagrams diverge quadratically, 
but the large N limit extracts planar one 
(planar limit). 

 We return to the side of noncommutative field theory. 
 There, the direct computation shows that 
\begin{eqnarray} 
 \Pi^{\rm NC}_{\rm planar}(p, \lambda) 
  &=& 
    \frac{\lambda}{2} 
    \int \frac{d^4 q}{(2\pi)^4} 
    \frac{1}{q^2 + m^2} \, , 
     \nonumber \\ 
  \Pi^{\rm NC}_{\rm nonplanar}(p, \lambda) 
  &=& 
    \frac{\lambda}{4} 
    \int \frac{d^4 q}{(2\pi)^4} 
    e^{i p \wedge q} 
    \frac{1}{q^2 + m^2} \, , 
\end{eqnarray} 
 We see that the planar diagram in Fig.~\ref{fig:two_point}(a) 
gets no phase factors. 
 Its contribution coincides with that of large N field theory 
\begin{equation} 
 \Pi^{\rm NC}_{\rm planar}(p, \lambda) = 
 \Pi^{\rm large\ N}_{\rm planar}(p, \lambda) \, , 
\end{equation} 
and diverges quadratically. 
 However, the nonplanar diagram in Fig.~\ref{fig:two_point}(b) 
gets nonzero phase factor. 
 We would like to see 
what is the effect of such a phase factor. 

 The Schwinger parametrization of the propagator 
enables us to perform the momentum integration 
\begin{eqnarray} 
 && 
 \Pi^{\rm NC}_{\rm nonplanar}(p, \lambda) 
 = \frac{\lambda}{4} \frac{1}{16\pi^2} 
    \int_0^\infty \frac{d \alpha}{\alpha^2} \, 
    \exp \left( 
          - \alpha m^2 - \frac{\widetilde{p}^2}{4} \frac{1}{\alpha} 
         \right) \, , 
   \label{eq:nonplanar} 
\end{eqnarray} 
where $\widetilde{p}^\mu = C^{\mu\nu} p_\nu$. 
 Then the UV limit is translated to the vanishing limit of $\alpha$. 
 Nonzero noncommutative parameter ensures that 
the integral converges 
since the exponentially suppression factor works 
when $\frac{1}{\alpha} \rightarrow \infty$. 
 The conclusion is that nonplanar diagram 
is UV-finite in noncommutative field theory. 

 Recalling that planar diagram contributions are 
the same in both theories, 
the UV limit of 
noncommutative field theory is equivalent to 
the UV and planar limit of 
the corresponding large N field theory 
\cite{Bigatti,Ishibashi_NCYM}. 
 This is the aspects of UV limit of noncommutative field theory. 
 It is determined by the planar diagrams. 

\section{IR aspect of noncommutative field theory} 
\label{sec:IR} 
 Next we would like to examine IR limit of 
noncommutative field theory.
 For that purpose, 
we return to the nonplanar contribution (\ref{eq:nonplanar}) 
to the two-point function. 
 The simple rescaling $\alpha = \widetilde{p}^2 t$ shows that 
this diverges in the IR side quadratically 
\begin{equation} 
 \Pi^{\rm NC}_{\rm nonplanar}(p, \lambda) 
  \propto \frac{1}{\widetilde{p}^2} 
  \quad {\rm for}\ p_\mu \rightarrow 0 
  \, . 
\end{equation} 
 To pursue its origin, 
we set $C^{\mu\nu}$ to zero. 
 Then, the integral is that of the planar diagram, 
which diverges quadratically. 
 It can be regularized by 
introducing the ultraviolet cut-off $\Lambda$ \cite{MRS} 
\begin{eqnarray} 
 && 
 \Pi^{\rm NC}_{\rm planar}(p, \lambda) 
 \propto 
 \frac{1}{16\pi^2} 
 \int_0^\infty \frac{d \alpha}{\alpha^2} \, 
  \exp 
  \left( 
    \alpha m^2 - \frac{1}{4} \frac{1}{\Lambda^2} \frac{1}{\alpha} 
  \right)\, . 
  \label{eq:planar} 
\end{eqnarray} 
 The cutoff-dependence found above 
reflects the quadratic divergence 
of the planar diagram. 
  Comparison of the planar contribution (\ref{eq:planar}) 
to the nonplanar contribution (\ref{eq:nonplanar}) shows 
that nonzero $C^{\mu\nu}$ and $p_\mu$ replace 
the cutoff dependence of the planar diagrams 
with $1/\widetilde{p}^2$ in nonplanar diagrams. 
 The IR-divergent behavior generated by the nonplanar diagrams 
reflects the UV-divergent behavior of the planar diagrams 
\cite{MRS,Hayakawa}. 

 We examine more interesting system, i.e, 
U(1) noncommutative Yang-Mills theory, in detail. 
 We consider the transverse part of 
the renormalized vacuum polarization for the photon 
\footnote{ 
 There arises 
another hard singular term 
proportional to the Lorentz structure, 
which is intrinsic to the nonvanishing noncommutativity 
$C^{\mu\nu}$. 
 Its implication is discussed in Ref. \cite{Matusis:NC}. 
}. 
 Its ultraviolet behavior is dominated by 
the planar diagrams 
\begin{eqnarray} 
 && 
 \left. 
  \Pi_{\mu\nu}(q) 
 \right|_{\rm transverse} 
 \rightarrow 
  - \frac{g^2}{16\pi^2} \frac{10}{3} {\rm ln}(q^2) 
    \left( 
     \delta_{\mu\nu} q^2 - q_\mu q_\nu 
    \right) 
    \quad 
     {\rm for}\ 
     \left| q \right| \gg \frac{1}{\sqrt{\left| C \right|}} 
     \, . 
      \nonumber \\ 
\end{eqnarray} 
 The infrared limit, 
which is now dominated by nonplanar diagrams. 
can be computed \cite{Hayakawa} 
\begin{eqnarray} 
 && 
 \left. 
  \Pi_{\mu\nu}(q) 
 \right|_{\rm transverse} 
 \rightarrow 
  - \frac{g^2}{16\pi^2} \frac{10}{3} 
    \left( -{\rm ln}(\tilde{q}^2) \right) 
    \left( 
     \delta_{\mu\nu} q^2 - q_\mu q_\nu 
    \right) 
    \quad 
    {\rm for}\ 
     \left| q \right| \ll \frac{1}{\sqrt{\left| C \right|}} 
     \, . 
      \nonumber \\ 
\end{eqnarray} 
 Note that $\left( -{\rm ln}(\tilde{q}^2) \right)$ 
is positive for $\left| q \right| \ll 1/\sqrt{\left| C \right|}$. 
 From those results, 
we see that 
the logarithmic nature of singularities 
coincide with each other. 
 Furthermore, both limiting behaviors coincide with each other,
including a numerical coefficient 
\footnote{
 This relation has been argued 
from point of view of string world-sheet. 
 See Ref.~\cite{string_consideration} and the references therein}. 


 There is an example which 
does not give such a precise correspondence between the infrared 
and ultraviolet sides as found in NCYM theory. 
 It is noncommutative QED, 
which is a noncommutative generalization of QED 
\cite{Hayakawa}. 
 Another aspect of that theory is that 
we cannot find its counterpart of large N field theory 
associated with noncommutative QED 
\footnote{ 
${\rm U}(N_C)$ NCYM theory with the number $N_F$ of quarks 
has its counterpart of large N field theory, 
i.e., ${\rm SU}(N_C)$ QCD with $N_F$ quarks 
\cite{Armoni}. 
}. 

\section{Conclusion and discussion}
\label{sec:conclusion} 

 Here we have observed a few fundamental properties of 
noncommutative quantum field theory. 
 Its UV limit is governed by planar diagrams, 
and usually also described by the corresponding 
large N field theory. 

 We have also seen that a new type of 
singularity in the infrared side is generated 
by the nonplanar diagrams, 
and it has a close relationship to the behavior at UV limit. 

 It is interesting to ask 
the practical issue 
whether noncommutative quantum field theory 
accommodates the quantum theory of gravity and string 
(See the recent attempts in Ref.~\cite{NP_NC}), 
especially 
whether the connection of IR and UV sides 
observed in NCYM theory is manifestation 
of some duality nature (e.g. closed-open channel duality) 
of string theory. 

\section*{Acknowledgements} 
 The author thanks to financial support 
from theory group at KEK.


\begin{thebibliography}{10} 
%
\bibitem{Bigatti} 
 D. Bigatti and L. Susskind, 
  {\it Magnetic fields, branes and noncommutative geometry}, 
   hep-th/9908056. 
%
\bibitem{Ishibashi_NCYM} 
 N. Ishibashi, S. Iso, H. Kawai and Y. Kitazawa, 
  {\it Wilson loops in noncommutative Yang-Mills}, 
   Nucl.\ Phys.\ {\bf B573}, 573 (2000), 
   hep-th/9910004. 
%
\bibitem{Connes} 
 M. Li, 
  {\it `Strings from IIB matrices}, 
  Nucl.\ Phys.\ {\bf B499}, 149 (1997), 
  hep-th/9612222; \\ 
 A. Connes, M. R. Douglas and A. Schwarz, 
  {\it Noncommutative geometry and matrix theory: 
       Compactification on tori}, 
  JHEP\ {\bf 9802}, 003 (1998), 
  hep-th/9711162. 
%
 \bibitem{Aoki_NCYM} 
 H. Aoki, N. Ishibashi, S. Iso, H. Kawai, 
 Y. Kitazawa and T. Tada, 
   {\it Noncommutative Yang-Mills in IIB matrix model}, 
   Nucl.\ Phys.\ {\bf B565}, 176 (2000), 
   hep-th/9908141. 
%
 \bibitem{BFSS} 
  T. Banks, W. Fischler, S. H. Shenker and L. Susskind, 
    {\it M theory as a matrix model: a conjecture}, 
    Phys.\ Rev.\  {\bf D55}, 5112 (1997), 
    hep-th/9610043. 
%
 \bibitem{IKKT} 
 V. Periwal, 
   {\it Matrices on a point as the theory of everything}, 
    Phys.\ Rev.\ {\bf D55}, 1711 (1997), 
    hep-th/9611103; \\ 
 N.~Ishibashi, H.~Kawai, Y.~Kitazawa and A.~Tsuchiya, 
   {\it A large-N reduced model as superstring}, 
    Nucl.\ Phys.\ {\bf B498}, 467 (1997), 
    hep-th/9612115. 
%
 \bibitem{DVV} 
  L.~Motl, 
   {\it Proposals on nonperturbative 
        superstring interactions}, 
   hep-th/9701025; \\ 
  R.~Dijkgraaf, E.~Verlinde and H.~Verlinde, 
   {\it Matrix string theory}, 
    Nucl.\ Phys.\  {\bf B500}, 43 (1997), 
    hep-th/9703030. 
%
 \bibitem{Zacho} 
  C.~Zachos, 
  {\it A survey of star product geometry}, 
  hep-th/0008010. 
%
 \bibitem{Seiberg} 
  N.~Seiberg, 
  {\it A note on background independence 
    in noncommutative gauge theories, 
    matrix model and tachyon condensation}, 
  JHEP\ {\bf 0009}, 003 (2000), 
  hep-th/0008013. 
%
 \bibitem{Pre_NC} 
  J.~C.~Varilly and J.~M.~Gracia-Bondia, 
  {\it On the ultraviolet behaviour of quantum fields over 
    noncommutative  manifolds}, 
   Int.~J.~ Mod.~Phys.~{\bf A14}, 1305 (1999), 
   hep-th/9804001; \\ 
  M.~Chaichian, A.~Demichev and P.~Presnajder, 
  {\it Quantum field theory on noncommutative space-times 
    and the persistence of ultraviolet divergences}, 
   hep-th/9812180; \\ 
  C. P. Martin and D. Sanchez-Ruiz, 
  {\it The one-loop UV divergent structure of 
    U(1) Yang-Mills theory on noncommutative $R^4$}, 
    Phys.~Rev.~Lett.~{\bf 83}, 476 (1999), 
    hep-th/9903077. 
  M. M. Sheikh-Jabbari, 
  {\it One loop renormalizability of supersymmetric Yang-Mills 
    theories on noncommutative two-torus}, 
    JHEP~{\bf 06}, 015 (1999), hep-th/9903107; \\ 
  T. Krajewski and R. Wulkenhaar, 
  {\it Perturbative quantum gauge fields 
    on the noncommutative torus}, 
    hep-th/9903187; \\ 
  M.~Chaichian, A.~Demichev and P.~Presnajder, 
  {\it Quantum field theory on the noncommutative plane 
    with E(q)(2) symmetry}, 
    hep-th/9904132; \\ 
  I.~Chepelev and R.~Roiban, 
  {\it Renormalization of quantum field theories 
    on noncommutative $R^d$. I: Scalars}, 
    hep-th/9911098; \\ 
  {\it Convergence theorem for non-commutative 
    Feynman graphs and renormalization}, 
    hep-th/0008090; \\ 
  I. Ya. Aref'eva, D. M. Belov and A. S. Koshelev,
  {\it Two-loop diagrams in noncommutative $\phi^4_4$ theory}, 
    hep-th/9912075. 
%
 \bibitem{MRS} 
  S. Minwalla, M. V. Raamsdonk and N. Seiberg, 
   {\it Noncommutative perturbative dynamics}, 
     hep-th/9912072. 
%
 \bibitem{Hayakawa} 
  M.~Hayakawa, 
   {\it Perturbative analysis on infrared aspects 
        of noncommutative QED on  $R^4$}, 
    {\it Phys.\ Lett.}\ {\bf B478}, 394 (2000), 
    hep-th/9912094; \\ 
   {\it Perturbative analysis on infrared and ultraviolet aspects 
        of noncommutative QED on $R^4$}, 
    hep-th/9912167. 
%
 \bibitem{'tHooft:planar} 
  G.~'t Hooft, 
    {\it A Planar Diagram Theory For Strong Interactions}, 
    Nucl.\ Phys.\  {\bf B72}, 461 (1974). 
%
 \bibitem{Matusis:NC} 
  A.~Matusis, L.~Susskind and N.~Toumbas, 
  {\it The IR/UV connection in the non-commutative 
    gauge theories}, 
  hep-th/0002075. 
%
 \bibitem{string_consideration} 
  J.~Gomis, M.~Kleban, T.~Mehen, M.~Rangamani and S.~Shenker, 
   {\it Noncommutative gauge dynamics from the string worldsheet}, 
   JHEP {\bf 0008}, 011 (2000), 
   hep-th/0003215; \\ 
  H.~Liu and J.~Michelson, 
   {\it Stretched strings in noncommutative field theory}, 
   Phys.\ Rev.\ {\bf D62}, 066003 (2000), 
   hep-th/0004013; \\ 
  S.~Kar, 
   {\it D-branes, cyclic symmetry and noncommutative geometry}, 
   hep-th/0006073. 
%
 \bibitem{Armoni} 
 A.~Armoni, 
  {\it Comments on perturbative dynamics of 
      non-commutative Yang-Mills theory}, 
  hep-th/0005208.
%
 \bibitem{NP_NC} 
  N.~Ishibashi, S.~Iso, H.~Kawai and Y.~Kitazawa, 
  {\it String scale in noncommutative Yang-Mills}, 
   Nucl.\ Phys.\  {\bf B583}, 159 (2000), 
   hep-th/0004038; \\ 
  S.~R.~Das and S.~Rey, 
  {\it Open Wilson lines in noncommutative gauge theory 
       and tomography of holographic dual supergravity}, 
   hep-th/0008042 \\ 
  D.~J.~Gross, A.~Hashimoto and N.~Itzhaki, 
   {\it Observables of non-commutative gauge theories}, 
   hep-th/0008075. 
\end{thebibliography}
\end{document}